\documentclass[aps,pre,twocolumn,superscriptaddress,showpacs]{revtex4}

\usepackage{graphicx}

\usepackage{dcolumn}

\begin{document}


\title{Roughening, deroughening, and nonuniversal scaling of the interface width in electrophoretic deposition of polymer chains}


\author{Frank W. Bentrem}
\affiliation{Program in Scientific Computing, The University of Southern Mississippi, Hattiesburg, Mississippi 39406-5046}

\author{Ras Pandey}

\affiliation{Department of Physics and Astronomy, The University of Southern Mississippi, Hattiesburg, Mississippi 39406-5046}

\author{Fereydoon Family}

\affiliation{Department of Physics, Emory University, Atlanta, Georgia 30322}


\date{July 2000}

\begin{abstract}

Growth and roughness of the interface of deposited polymer chains
driven by a field onto an impenetrable adsorbing surface are studied by
computer simulations in ($2+1$) dimensions. The evolution of the
interface width $W$ shows a crossover from short-time growth described
by the exponent $\beta_1$ to a long-time growth with exponent $\beta_2$ ($>\beta_1$). The
saturated width increases, i.e., the interface roughens, with the
molecular weight $L_c$, but the roughness exponent $\alpha$ (from $W_s \sim L^\alpha$) becomes
negative in contrast to models for particle deposition; depends on the
chain length-—-a nonuniversal scaling with the substrate length
$L$. Roughening and deroughening occur as the field $E$ and the temperature
$T$ compete such that $W_s \approx (A+BT)E^{-1/2}$.

\end{abstract}


\maketitle


As polymer chains are driven by a field toward a substrate, e.g., the
pore boundary of a gel, in a DNA electrophoresis \cite{lerman82,lumpkin85,slater89,hoagland96,quake97,perkins97}, it is not clear
how the polymer interface width $W$ evolves and scales with multiscaling
fields such as molecular weight $L_c$ (chain length), electric field $E$, and
temperature $T$. Understanding of the driven polymer interface is also
important in the design of composite materials via molecular
deposition and evolution of their physical properties \cite{tsao93,palasantzas98,fleer93}. In
coatings technology \cite{brewer73}, knowledge of the density profile and roughness
of the polymer layer on the adsorbing surface \cite{fleer93,pandey97} is the key to
characterizing surface properties \cite{foo98}. In general, these studies are
useful in understanding polymer characteristics at the interface \cite{deconinck95} in
mixtures or adhesion of dissimilar substances. 

While considerable
progress has emerged in the understanding of the growth of the
interface width and its scaling in deposition of particles \cite{family91,yang93,barabasi95}, very
little is known about the interfacial dynamics in polymer deposition
\cite{foo98,collins94,wool95,foo99} . We find unexpected results for the growth of the driven
polymer interface width, which exhibits unusual nonuniversal scaling
with multiscaling variables ($L_c$, $E$, $T$). In the following, we show
that the interface width (i) decreases, i.e., deroughens, with
increasing field strength and (ii) increases with increasing temperature
and molecular weight, but the magnitude of the deroughening exponent
is enhanced with a larger molecular weight. Reduction in surface
roughness by the polymer chains in the deposition process may be
related to drag reduction by polymers in fluid flow \cite{hoyt91}. 

\begin{figure}

\includegraphics[width=3.34in]{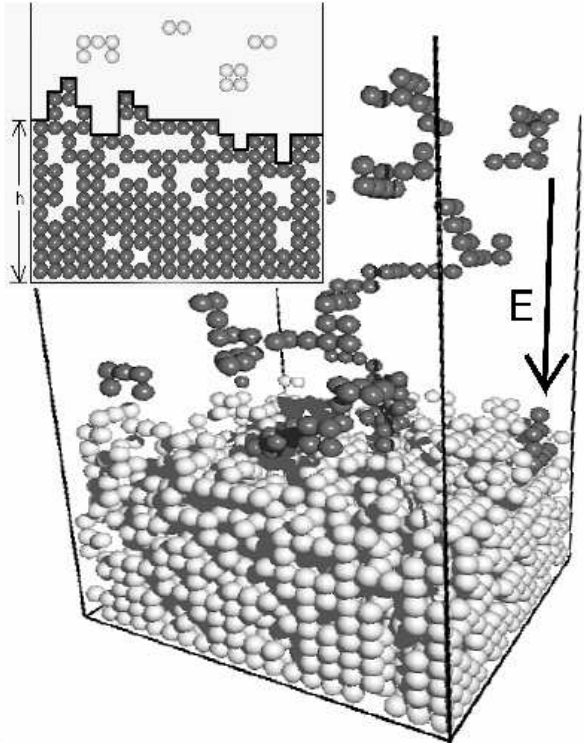}

\caption{Snapshot of a $40 \times 20 \times 20$ system with chains of
length $L_c=10$ at $E=0.5$ and $T=1$. The light beads in bulk are
polymer nodes that are connected to the wall via a network of occupied
nearest-neighbor sites. The charged polymer chains are driven from top
to bottom ($x$ direction) by a uniform electric field $E$. The inset
shows a two-dimensional slice in the $xy$ plane. The height $h_{ij}$ of
the interface at each column is the distance from the substrate (bottom)
to the highest dark bead in that column.}

\label{fig1}

\end{figure}

Using Monte
Carlo simulations, we study the surface dynamics of polymer chains
driven by a field toward an impenetrable adsorbing barrier in a ($2+
1$)-dimensional discrete lattice. We consider polymer chains, each of
length $L_c$ , which consist of $L_c+1$ nodes connected consecutively by
rigid unit bonds along the lattice using a self-avoiding-walk
constraint. All lengths (chain, substrate) are measured in arbitrary
units of the lattice constant. In polymer simulations \cite{binder95},
implementation of the local dynamics of polymer chain movement is very
important in studying the physical properties on various time
scales. Kink-jump dynamics is used according to the Verdier-Stockmayer
algorithm \cite{verdier62} to move chain nodes on a $L_x \times L \times L$ lattice with a large
aspect ratio $L_x/L$. It should be pointed out that the kink-jump move
provides a slow chain dynamics in contrast to the relatively faster
dynamics used in a recent study \cite{foo98} where a reptation move is
implemented in combination with the kink-jump moves. Some of the
microscopic details arising from the slow kink-jump moves are
generally lost when reptation is implemented with the kink jump since
the former dominates over the latter. The details of conformational
and density evolution are, therefore, better taken into account due to
small scale kink-jump dynamics for the model considered here. Chains
are released at nearly a constant rate from a source end (near $x=1$) of
the sample and are driven by an electric field toward an impenetrable
adsorbing wall (in the $yz$ plane) at $x=L_x$. In addition to excluded
volume, nearest-neighbor polymer-polymer repulsive and polymer-wall
attractive adsorbing interactions are implemented. The external field $E$
couples with the change in energy, $\delta E=Edx$, for the displacement $dx=0$ or
$\pm1$ of each node along the $x$ direction. A randomly selected chain node
is attempted to move according to the Metropolis algorithm \cite{metropolis53}. $N$
attempts to move randomly selected nodes are defined as one Monte Carlo
step (MCS), where $N$ is the number of occupied lattice sites. Even
though the measure of time in units of MCS is arbitrary, it provides a
way to analyze the growth and evolution of quantities such as
interface width with the MCS time, i.e., $W \sim t^\beta$. However, it is observed
that the interface width reaches steady state, i.e., saturates, and so
the scaling of the saturated width $W_s$ with the substrate length $W_s \sim L^\alpha$, where $\alpha$ is the roughness exponent, and with other parameters such as
field and temperature (see below) is independent of the time unit. A
snapshot of the system is shown in Fig. \ref{fig1}. 

\begin{figure}

\includegraphics[width=3.34in]{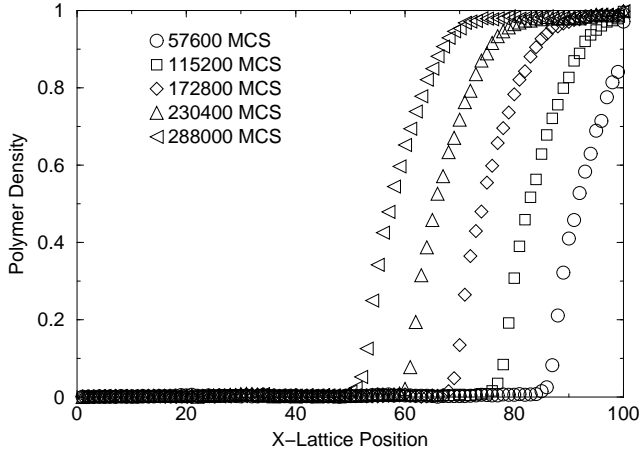}

\caption{Density profile of polymer coating on a $100 \times 30 \times
30$ lattice for polymer chains of length $L_c=30$ at $T=1$ and
$E=0.5$. Plots are shown for several different times in units of Monte
Carlo time steps (MCS). The $x$ lattice position is measured in units
of the lattice constant, and polymer density is the fraction of
occupied sites.}

\label{fig2}

\end{figure}

Figure \ref{fig2} shows a typical evolution of the polymer density
profile. Because of a high degree of ramification, it is rather
difficult to identify the surface; therefore, an accurate algorithm is
needed with good statistics to evaluate the average height $\overline
h$ of the growing surface and its fluctuation, i.e., the interface
width

\begin{equation}
W^2 = \frac{\sum\limits_{ij}(h_{ij}-\overline h)^2}{L^2},
\end{equation}

\noindent
where

\begin{equation}
\overline h = \frac{\sum\limits_{ij}h_{ij}}{L^2},
\end{equation}

\begin{figure}

\includegraphics[width=3.34in]{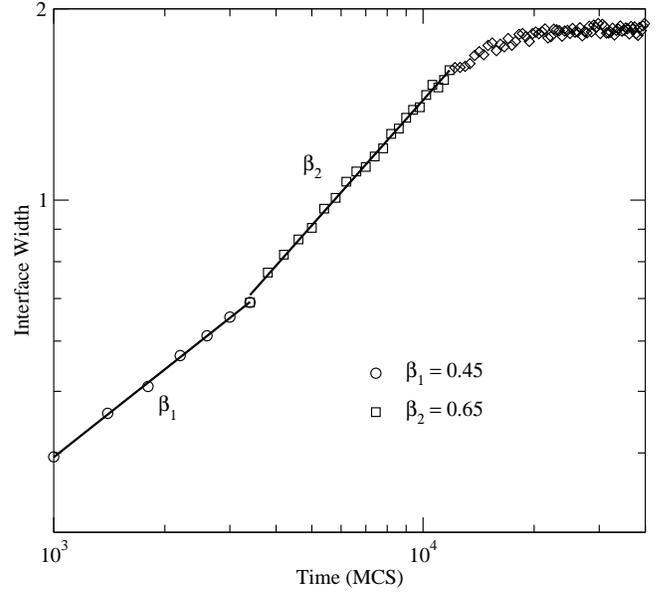}

\caption{Growth of the interface width $W$ in units of the lattice
constant with time on a log-log scale for $L=30$ and $L_c=5$ at
$E=0.5$ and $T=1$.}

\label{fig3}

\end{figure}

\noindent
and $h_{ij}$ is the surface height at location ($i,j$) on the
wall/substrate. We identify the polymer surface coating as the locus
of polymer nodes where each polymer node is connected to the adsorbing
wall with a nearest-neighbor network of nodes as in cluster
identification in standard percolation \cite{stauffer94}. Evolution of
a typical interface width is presented in Fig. 3. We note that the
interface width initially grows comparatively slowly in the short-time
regime ($t \leq t_1$) followed by a faster-growth regime ($t_1 \leq t
\leq t_2$) before reaching saturation in the longtime regime ($t \gg
t_2$). We examine the growth of the interface width in these regimes
as power laws, i.e., $W \sim t^{\beta_1}$ in $t \leq t_1$ and $W \sim
t_2$ in $t_1 \leq t \leq t_2$. Estimates of these growth exponents for
various values of chain length and field are presented in Tables I and
II, respectively. We find that $\beta_1 \leq \beta_2$ and the growth
exponents have the values $\beta_1 \approx 0.50 \pm 0.10$ and $\beta_2
\approx 0.66 \pm 0.10$ for $E=0.5$ and $T=1$ and are nearly
independent of chain length. The higher values of the growth exponent
$\beta_2$ seem consistent with the experimental measurements of the
surface roughness of plasma polymer films \cite{collins94}. We note
that the magnitude of the growth exponents in both initial and
intermediate growth regimes (i.e., 0-–-$t_1$, $t_1$–--$t_2$) depend on the
strength of the field, and therefore, the growth exponents are
nonuniversal. 

\begin{table}

\caption{Growth exponents $\beta_1$
(initial) and $\beta_2$ (before saturation) with substrate length $L=30$ at $E=0.5$, $T=1$ for different chain lengths.}

\begin{ruledtabular}

\begin{tabular}{ccc}

\hbox{Chain length } $L_c$ & $\beta_1$ & $\beta_2$ \\

\hline

2 & $0.46\pm0.01$ & $0.64\pm0.01$ \\

3 & $0.51\pm0.01$ & $0.60\pm0.02$ \\

4 & $0.44\pm0.01$ & $0.63\pm0.01$ \\

5 & $0.45\pm0.01$ & $0.65\pm0.01$ \\

9 & $0.41\pm0.03$ & $0.69\pm0.01$ \\

11 & $0.43\pm0.05$ & $0.69\pm0.02$ \\

14 & $0.56\pm0.05$ & $0.72\pm0.03$ \\

19 & $0.41\pm0.07$ & $0.50\pm0.03$ \\

25 & $0.41\pm0.04$ & $0.57\pm0.03$ \\

\end{tabular}

\end{ruledtabular}

\end{table}

\begin{table}

\caption{Growth exponents $\beta_1$ and $\beta_2$ with $L=30$ and
$L_c=25$ at $T=1$ for different field strengths.}

\begin{ruledtabular}

\begin{tabular}{ccc}

Field $E$ & $\beta_1$ & $\beta_2$ \\

\hline

0.08 & $1.05\pm0.07$ & $1.08\pm0.04$ \\

0.09 & $0.88\pm0.05$ & $1.05\pm0.07$ \\

0.10 & $0.97\pm0.10$ & $1.13\pm0.04$ \\

0.20 & $0.72\pm0.08$ & $0.76\pm0.06$ \\

0.30 & $0.49\pm0.04$ & $0.62\pm0.05$ \\

0.40 & $0.36\pm0.03$ & $0.55\pm0.06$ \\

0.50 & $0.41\pm0.04$ & $0.57\pm0.03$ \\

0.60 & $0.40\pm0.04$ & $0.47\pm0.04$ \\

0.70 & $0.30\pm0.05$ & $0.48\pm0.14$ \\

\end{tabular}

\end{ruledtabular}

\end{table}

\begin{figure}

\includegraphics[width=3.34in]{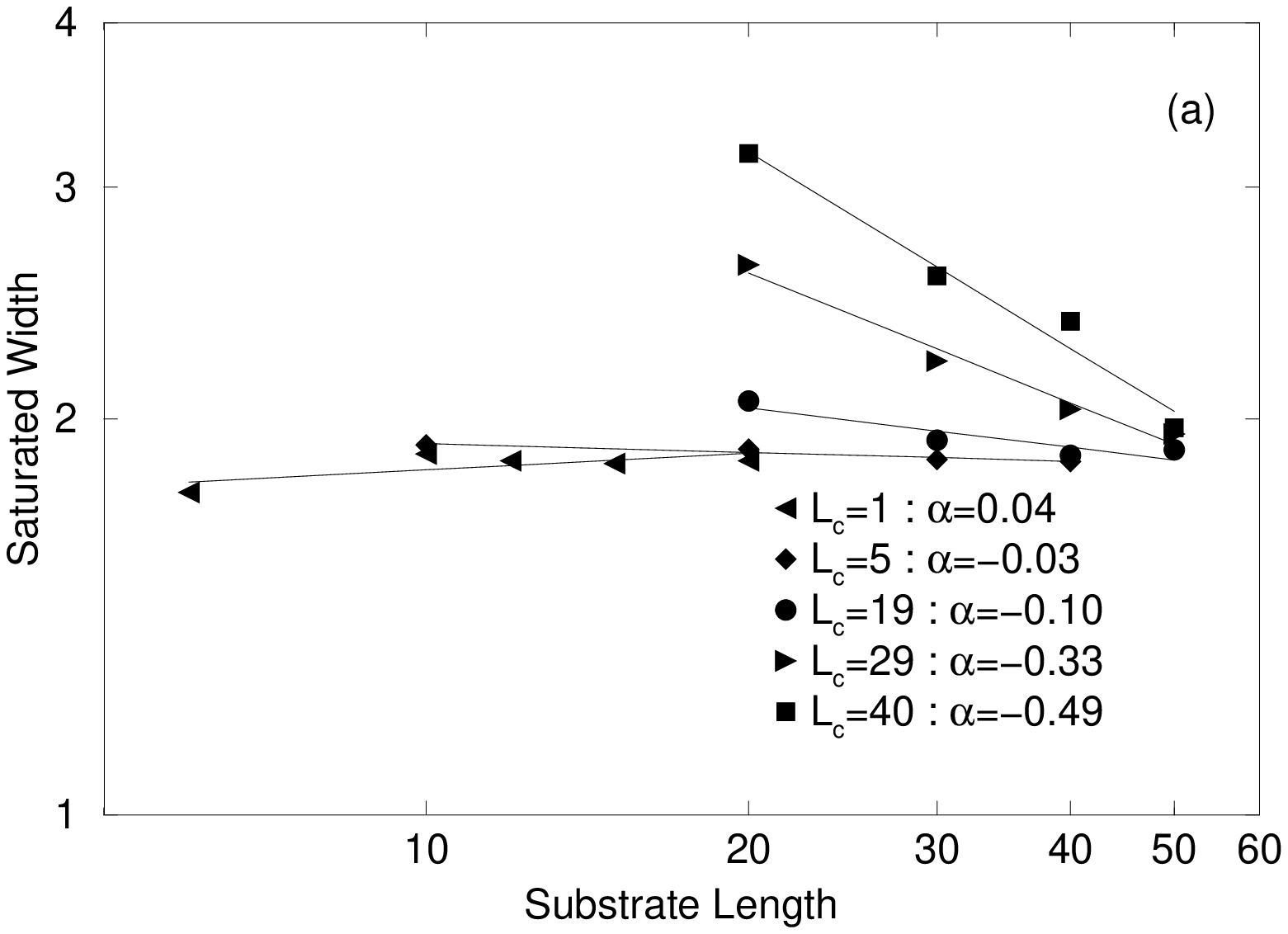}

\includegraphics[width=3.34in]{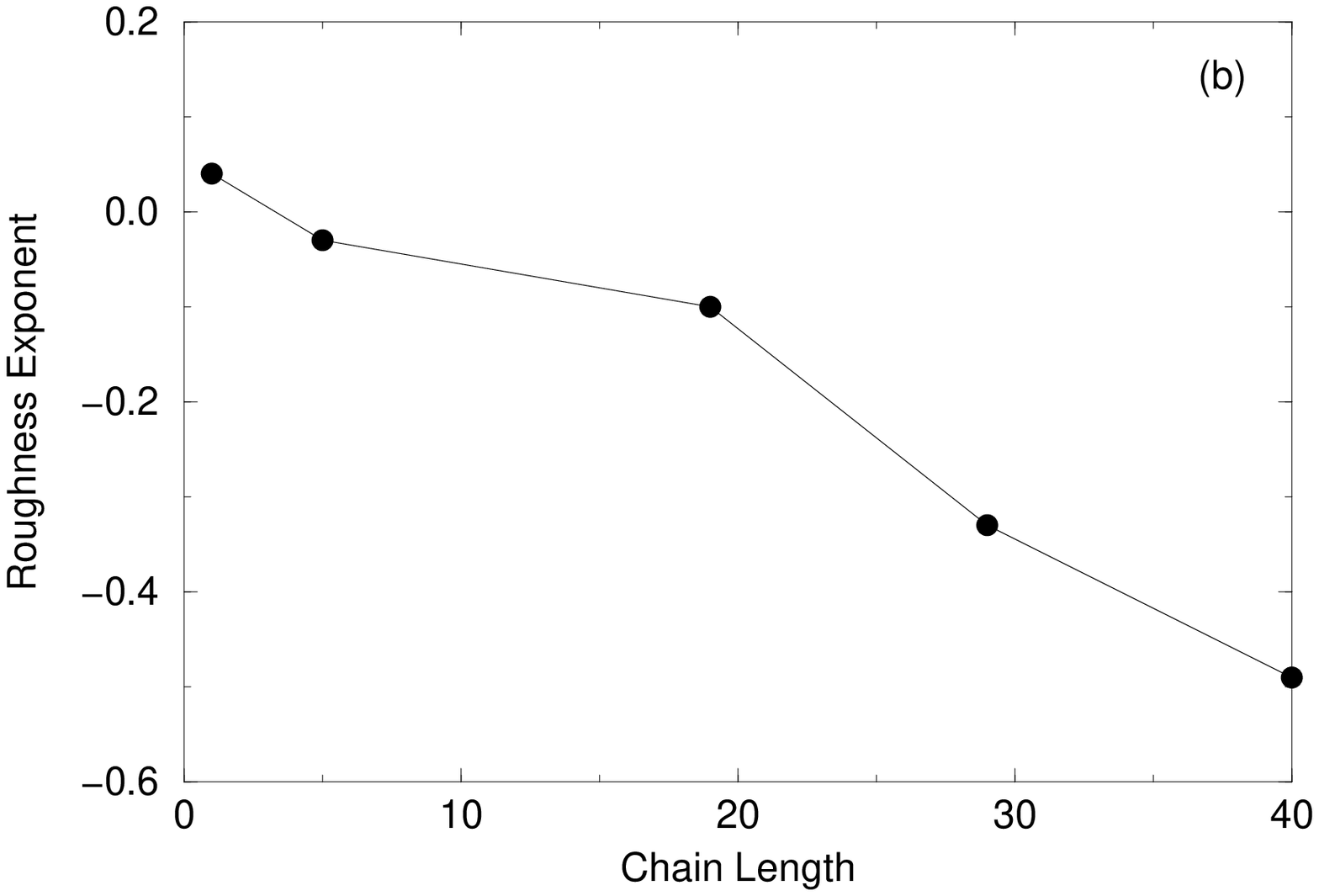}

\caption{(a) Saturated width versus substrate length (in units of the
lattice constant) on a log-log scale for $L_c=1$, 5, 19, 29, 40 at $T=1$
and $E=0.5$. (b) Variation of the roughness exponent with the chain length
(in units of the lattice constant).}

\label{fig4}

\end{figure}

Variation of the asymptotic or saturated width $W_s$ with $L_c$, $E$,
and $T$ is examined in detail. In the dynamic scaling analysis of
fluctuating surfaces \cite{family85}, one usually looks at the
finite-size scaling of the saturated width with the substrate
length $L$ to evaluate the roughness exponent $\alpha$ (i.e., $W_s
\sim L^\alpha$). Single-parameter scaling of the asymptotic width with
the substrate length L has been verified in detail via extensive
particle simulations \cite{family91,yang93,barabasi95}. In our simulations (see Fig. 4), the saturated
interface width is insensitive to the substrate length $L$ with $\alpha \approx 0$ for short chain lengths ($L_c=1$---5). With the longer chain lengths ($L_c \ge 10$),
on the other hand, $\alpha$ becomes negative, i.e., $W_s$ decreases with $L$ with a
negative value of the roughness exponent . To our knowledge, the
dependence of the roughness exponent on the molecular weight has not
been reported by computer simulation before, and the negative value of
the exponent with higher molecular weight is a finding we report
here. This implies that the roughness reduces upon increasing the
substrate length as the polymer chains access more space to relax
entropically. The (negative) magnitude of increases with the chain
length. (See the variation of $\alpha$ with the chain length in Fig. 4.) It is
important to point out that the phenomena of deroughening with the
substrate length $L$ observed here is opposite to most roughening
studies with particle deposition \cite{family91,yang93,barabasi95} and experiments on plasma polymer
film \cite{collins94}. At a fixed substrate size, we find that the saturated width
increases monotonically with the molecular weight $L_c$. 

\begin{figure}

\includegraphics[width=3.34in]{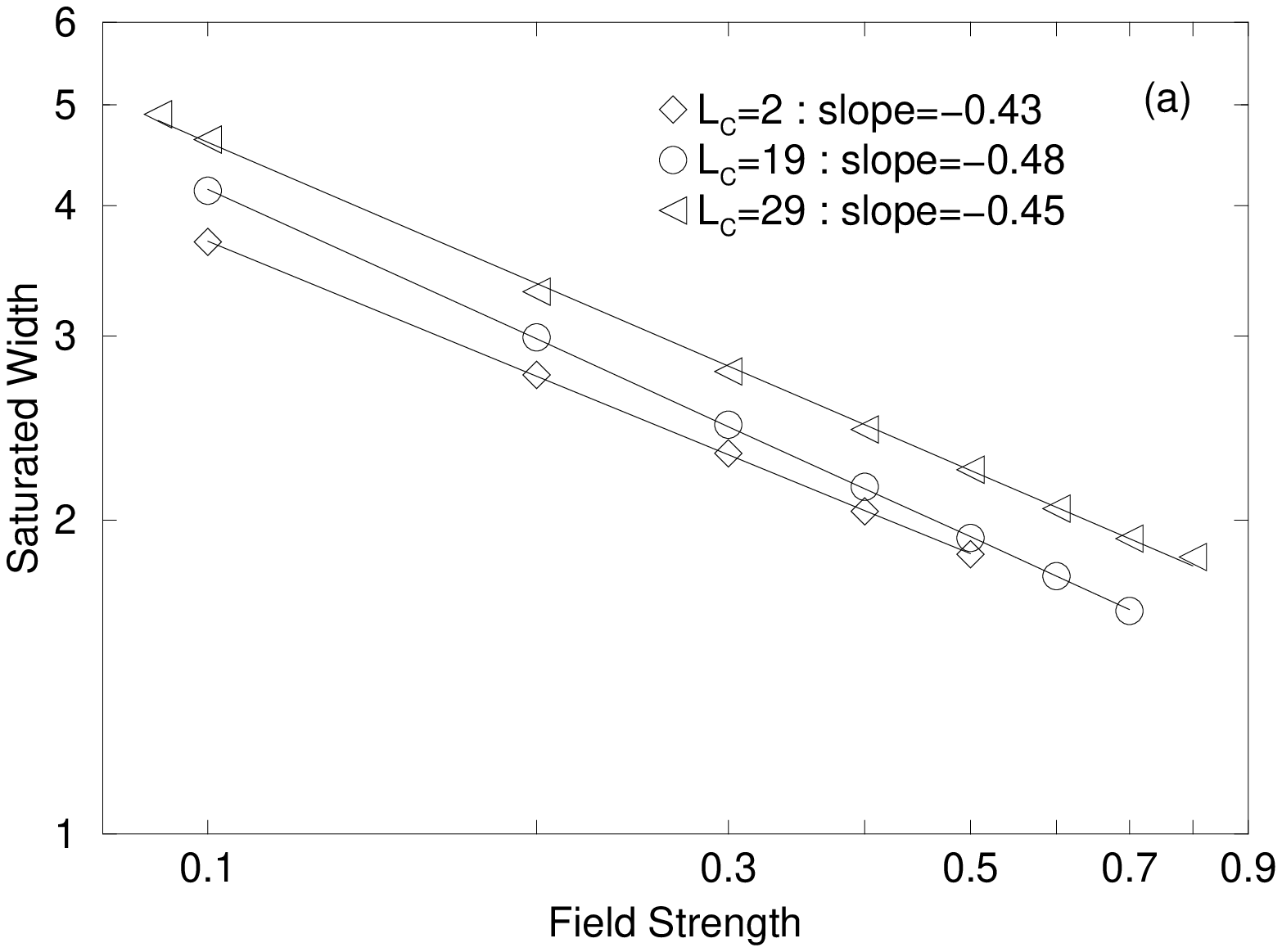}

\includegraphics[width=3.34in]{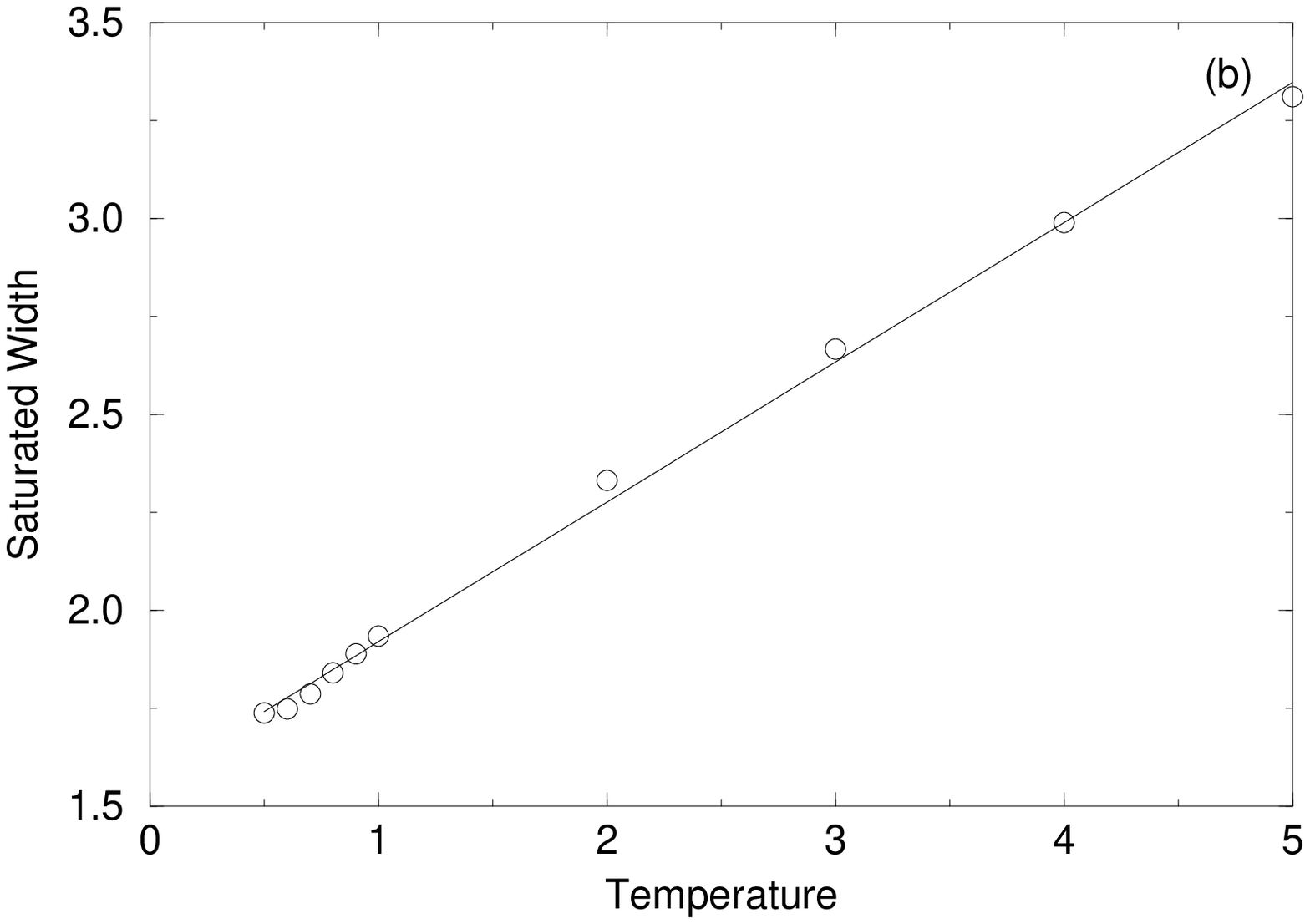}

\caption{Variation of saturated width (a) with field strength on a
log-log scale for $L=30$, $L_c=2$, 19, 29, and $T=1$, and (b) with
temperature for $L_c=19$, and $E=0.5$. Field and temperature are in
arbitrary units.}

\label{fig5}

\end{figure}

We have also examined the variation of the saturated width with the
field strength and temperature (see Fig. 5). We find that $W_s \sim
E^{-1/2}$ at the temperature $T=1$. Note that the saturation width
increases linearly with temperature unlike the plasma polymer
experiment \cite{collins94} where the interface width seems to
decrease with the temperature. In particle deposition models
\cite{family91,yang93,barabasi95}, one would expect a decrease in the
interface width with the temperature as the particles diffuse to
reduce the roughness. In our polymer system, on the other hand, each
polymer chain is uniformly charged so the monomer-monomer repulsion
causes the chains to take on an extended conformation. Upon contact
with the deposited surface, the field tends to cause the extended
chains to lie in the $yz$ plane leading to a small interface width. At
higher temperatures, however, the chains take on a more compact
random-coil conformation that yields a larger interface width. There
are various characteristic lengths involved in interfacial polymer
dynamics, e.g., the height-height correlation length $\xi_h$, radius of
gyration of chains $R_g$ , and density-density correlation length $\xi_d$,
which are all interdependent. Effects of external parameters ($E,T$) on
these lengths at the interface are difficult to evaluate. It is,
therefore, rather difficult to propose a multiparameter scaling for the
interfacial dynamics at this stage. However, it is clear that the
interface width depends on the field and the temperature described by
our empirical law $W_s \approx (A+BT)E^{-1/2}$, where $A$ and $B$ are increasing
constants. The observation that $W_s \sim L^{|\alpha|}$ with increasing magnitude of $|\alpha|$ with the chain length suggests
that the deroughening exponent ($-\alpha$) is nonuniversal.

 The
authors would like to thank A.-L. Barab\'{a}si, G. M. Foo, and M. Kardar
for useful discussions. We acknowledge help from G. M. Foo with the
computer program at the initial stage. This work was partially
supported by NSF Grant No. DMR-9520842 and by grants from NSF-EPSCoR,
DOEEPSCoR, the Office of Naval Research, and NASA. Visualization
performed at the High Performance Visualization Center (HPVC) was
helpful.

\bibliography{pre00}

\end{document}